\documentclass[aps,twocolumn,superscriptaddress,amsmath,amsfonts,
               floatfix]{revtex4}

\usepackage{epsfig,psfrag}
\usepackage{CJK}

\usepackage{graphicx}
\usepackage{dcolumn}
\usepackage{bm}
\usepackage{amssymb}
\usepackage{natbib}

\usepackage{setspace}

\begin{document}
\title{Phonon interference induced by defect pairs}
\author{Ran-Bo Yang}
\affiliation{Tianjin Key Laboratory of Low Dimensional Materials Physics and Preparing Technology, Department of Physics, School of Science, Tianjin University, Tianjin 300354 China}
\author{Ming-Hao Li}
\affiliation{Tianjin Key Laboratory of Low Dimensional Materials Physics and Preparing Technology, Department of Physics, School of Science, Tianjin University, Tianjin 300354 China}
\author{Chun-Yu Cai}
\affiliation{Tianjin Key Laboratory of Low Dimensional Materials Physics and Preparing Technology, Department of Physics, School of Science, Tianjin University, Tianjin 300354 China}
\author{Yue-Wen Wang}
\affiliation{Tianjin Key Laboratory of Low Dimensional Materials Physics and Preparing Technology, Department of Physics, School of Science, Tianjin University, Tianjin 300354 China}
\author{Zhi-Qing Li}
\affiliation{Tianjin Key Laboratory of Low Dimensional Materials Physics and Preparing Technology, Department of Physics, School of Science, Tianjin University, Tianjin 300354 China}
\author{Li-Xia Zhao$^{\dagger}$}
\affiliation{School of Electronics and Information Engineering, Tiangong University, Tianjin 300387, China}
\author{Zi-Wu Wang*}
\affiliation{Tianjin Key Laboratory of Low Dimensional Materials Physics and Preparing Technology, Department of Physics, School of Science, Tianjin University, Tianjin 300354 China}
\email{wangziwu@tju.edu.cn; lxzhao@tiangong.edu.cn}
\begin{abstract}
The coherent dynamics of lattice waves is of significant importance for exploring the nonadiabatic phenomena of the electron-phonon coupling under nonequilibrium conditions. Here, we propose a theoretical model for phonon interference induced by a defect pair, in which lattice waves with the selective modes are launched independently by each of defects due to the strongly local electron-phonon coupling. We find that the intensities of phonon interference fringes could be directly evaluated by a measurable quantity in experiments--Huang-Rhys factor, and the geometry of interference patterns could be modified by the internal parameters of defect pair, including the charge states, spatial distance, energy-level depth and localization strength. In particular, several novel quantum states, such as the Dirac-like and nodal-line topological states are emerged for the appropriate internal parameters. Furthermore, using Huang-Rhys model, we simulate the ultrafast optical spectroscopy in the phonon interference duration when defect pairs are excited as the color centers, where interference fringes are perfectly matched by intensities of spectrum, similar to the $\textit{lock-key}$ pattern. These results show that defect pair, serving as the driven force for the coherent dynamical processes for multidegrees of freedom, offers an ideal platform to explore fundamental quantum phenomena in nonequilibrium and ultrafast physics.
\end{abstract}

\maketitle

$\emph{Introduction}.$--Understanding of thermal or phonon transport in various electronic and photonic materials and devices is the cornerstone for effectively manipulating the processes of dissipate, store, or convert thermal energy\cite{y1,y2,j1,y3,y4}. In particular, the coherence and interference properties of phonons, associated with the wave-like properties of lattice vibration in few nanometers scale, determine the microscopic quantum processes of thermal conductivity\cite{y2,y4}. In the past decades, intense theoretical efforts for the phonon interference in the macroscopic solids\cite{y2,y4} and molecular systems\cite{y1,y3} and have been implemented in relatively large-scale artificial superlattices\cite{y5}, phonon crystals\cite{y6} and various nanostructure and molecular materials\cite{y1,y3}. More recently, ultrasensitive and stable scanning thermal microscope have allowed for the real-space detection of constructive and destructive phonon interference in the single molecular junction\cite{y7,y8}. However, the most severely challenge for phonon interference still lies in the distinguishing of detection signals for individual phonon mode from the multiple wavelength of thermally excited phonons and their intertwines.

The lattice vibration with the resolved single mode couples to the electronic and structural degrees of freedoms in materials, in fact, has been widely studied\cite{j2,e1,e2,e3,e4,e5,e6,e7,e8,e9,e10,e11,e12,e13,e14,e15,e16,e17,e18}, because this nonadiabatic dynamical process gives rise to a series of ultrafast phenomena under nonequilibrium conditions, such as the switching magnization\cite{e12}, the steering symmetry-breaking\cite{e4,e7}, the bright-dark states transitions\cite{e15} and transient topological band-structure renormalization\cite{e17,e18}. Different ultrafast optical spectroscopic technologies have been devoted to probe these nonadiabatic coupling effects, including femtosecond pump-probe spectroscopy\cite{e3,e4,e5,e6,e7}, time-resolved transient optical spectroscopy\cite{e9,e11,e12,e13,e14}, high-harmonic spectroscopy\cite{e16} and high-resolution resonant impulsive stimulated Raman spectroscopy\cite{e1,e8,e10}. Among them, the response signals of coherent lattice vibration are profiled effectively by the variations of optical intensities and tunneling currents. Several classical theoretical models\cite{t1,t2,t3,t4,t5} cooperating with first-principles calculations\cite{e15,e17,e18,t6} have been proposed for understanding the microscopic dynamical processes in these experiments and attributed to the mechanism of the displacive excitation of coherent phonon, launched by the optical field, electron-phonon and phonon-phonon interaction and even ultrafast Coulomb force\cite{t5}. Although the ultrafast dynamics of lattice waves with selected and resolved modes have been extensively investigated\cite{e15,e17,t6}, the completely quantum interference patterns for the displacive excitation of lattice waves, analogy to the interference effects of photons and electrons in temporal-spatial resolution, have yet not been presented.

Defects and impurities in solids regarded as localized centers, whose electronic states are influenced significantly by the equilibrium positions of nearby atoms, in turn, the positions of the neighboring atoms undergo substantial shifts when an electronic transition occurs, in which the local lattice waves are launched by these defect states due to the strong lattice distortion effect described by the well-known Huang-Rhys model\cite{h1,h2}. Here, we propose the interference of lattice vibrational waves driven by two independent defects in solid, analogous to two sources of water or sound waves, where the interference intensities of the constructive and destructive fringes could be directly profiled by an experimentally measurable quality--Huang-Rhys factor, allowing for depicting interference fringes of the selective lattice vibration modes in the real space. We find that the geometry of interference patterns could be substantially modulated by the distance, the ratios of energy-level depth and localization strength between two defects. In particular, the outlines of destructive fringes between two defects evolute from Dirac conical into the nodel-line semimetal states, displaying the novel topological states are induced by the electron-coherent phonon coupling. Furthermore, using Huang-Rhys model, we present the evolution of optical transition for defect pairs excited as color centers in the phonon interference durations, in which these interference fringes are excellently profiled by the reversal intensities of spectrum, suggesting that the phonon interference patterns could be discerned by ultrafast spectroscopic fingerprints.
\begin{figure}
\includegraphics[width=3.4in,keepaspectratio]{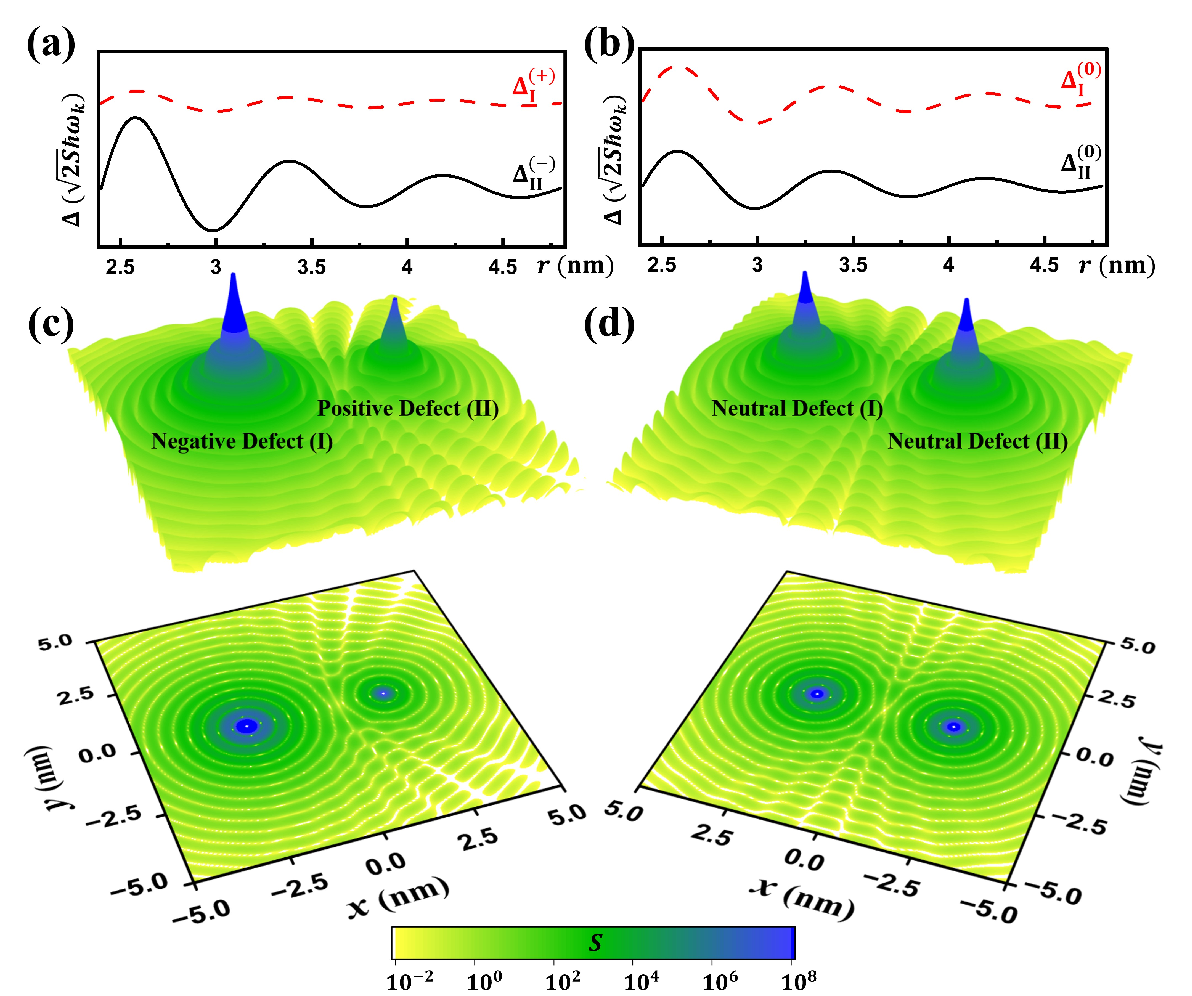}
\caption{\label{compare} The spatial evolution of the oscillation amplitudes of lattice waves induced by positive-negative defects pair (a) and neutral-neutral defect pair (b);  The corresponding quantum interference patterns of lattice waves for two types of defect pairs are plotted in (c) and (d), where the selected phonon frequency $\omega_k=$76 THz and the wave vector $k=$7.8 nm$^{-1}$ for the lattice waves with the fixed initial phase difference $\varphi_2-\varphi_1=0$. Here, the defect energy depth $E$=0.3 eV is adopted in the typical wide-bandgap semiconductor HfO$_2$ for two neutral defects as well as for the positive and negative defects respecting to the valence and conduction bands, respectively; The distance between two defects $D=4$ nm for two types of defect pairs with the same localization parameter.}
\end{figure}

$\emph{Huang-Rhys model for interference lattice waves.}$--
Defects or impurities in solid have the remarkably lattice distortion effect due to their local charge states, giving rise to the more stronger electron-phonon coupling than the normal electron states. In general, the displacement magnitude of lattice distortion out of equilibrium position for $k$th vibrational mode is described by a physical quantity $\Delta(k)$, which is related to the well-known Huang-Rhys factor\cite{h1,h2} by $S=\sum_k \Delta^2(k)/(\sqrt{2}\hbar\omega_k)^2$ ($\omega_k$ is the angular frequency for the $k$th phonon mode) that denotes the total lattice relaxation strength for different phonon modes, being able to evaluated directly in experiments\cite{h3,h4}. Here, we assume two independent defects in a traditional wide-gap semiconductor material HfO$_2$ (see the Table S1 in the Supplemental Materials), where the lattice waves with the fixed frequency and wave vector induced by them, respectively, can be expressed as
\begin{eqnarray}
\Delta_I(\textbf{k})&=& M(k)\rho_1(r_1)e^{i({\textbf{k}\cdot \textbf{r}_1}-\omega_k t+\varphi_1)}\nonumber\\
&=& A_1(k, r_1)e^{i({\textbf{k}\cdot \textbf{r}_1}-\omega_k t+\varphi_1)},
\end{eqnarray}
\begin{eqnarray}
\Delta_{II}(\textbf{k})&=&M(k)\rho_2(r_2)e^{i({\textbf{k}\cdot \textbf{r}_2}-\omega_k t+\varphi_2)}\nonumber\\
&=& A_2(k, r_2)e^{i({\textbf{k}\cdot \textbf{r}_2}-\omega_k t+\varphi_2)}.
\end{eqnarray}
Here, $M(k)$=$(\hbar \omega_k/k)\sqrt{4\pi\alpha/V}\sqrt[4]{\hbar/(2m^*\omega_k)}$ is the electron-optical phonon interaction matrix element in the Fr$\ddot{o}$hlich mechanism with the coupling constant $\alpha$, $m^*$ is the effective mass of electron (or hole) and $V$ is the volume of material. This coupling mechanism is selected because the coherence dynamics of optical phonon modes have been widely proved in most experiments\cite{j2,e1,e2,e9,e14,e16}. $\rho(r)$=$|\psi(r)|^2$ represents the charge density of defect, where a quantum defect model is adopted to describe the defect states\cite{d1,d2,d3}
\begin{equation}
\psi(r)=\frac{(\beta r/a^*)^{\mu-1}\exp[{-\beta r/(a^*\nu)}]}{\sqrt{4\pi a^{*3}(\nu/2)^{2\mu+1}\Gamma(2\mu+1)}},
\end{equation}
 with $a^*=4\pi\epsilon\hbar^2/(e^2m^*)$ is the effective Bohr radius and $\nu=e/({\sqrt {8\pi \epsilon{a^*E}}})$ with $e$ is the electronic charge, $\epsilon$ is the high-frequency permittivity, $E$ is the energy-level depth of defect in the bandgap and $\Gamma$ is the Gamma function. The charge state of defect is reflected by the parameter $\mu=+\nu, -\nu, 0$ standing for positive, negative and neutral defects, respectively. The localization properties of defects are reflected by two key parameters: (i) the energy-level depth of defect ($E$), the local lattice distortion (or called as the lattice relaxation effect) will be enhanced as the defect depth increases; (ii) $\beta$ denotes the localization parameter, assuming that the bigger $\beta$, the stronger localization for defects. Both of them determine the basic features for oscillation amplitude of lattice wave $A_i(k, r_i)$ ($i=1,2$). The evolution of lattice waves with these two parameters are illustrated in Fig. S1 of the Supplemental Materials.

Now, the superposition of lattice waves induced by two defects is expressed in term of Huang-Rhys factor as
\begin{equation}
S= \frac{|\Delta_{I}(\textbf{k})+\Delta_{II}(\textbf{k})|^2}{(\sqrt{2}\hbar\omega_k)^2},
\end{equation}
in which the summation for the phonon modes are dropped since the interference processes are inherently only for certain selected phonon mode. The detailed expressions for this equation are given in the Supplemental Materials.
\begin{figure*}
\centering
\includegraphics[width=6.8in,keepaspectratio]{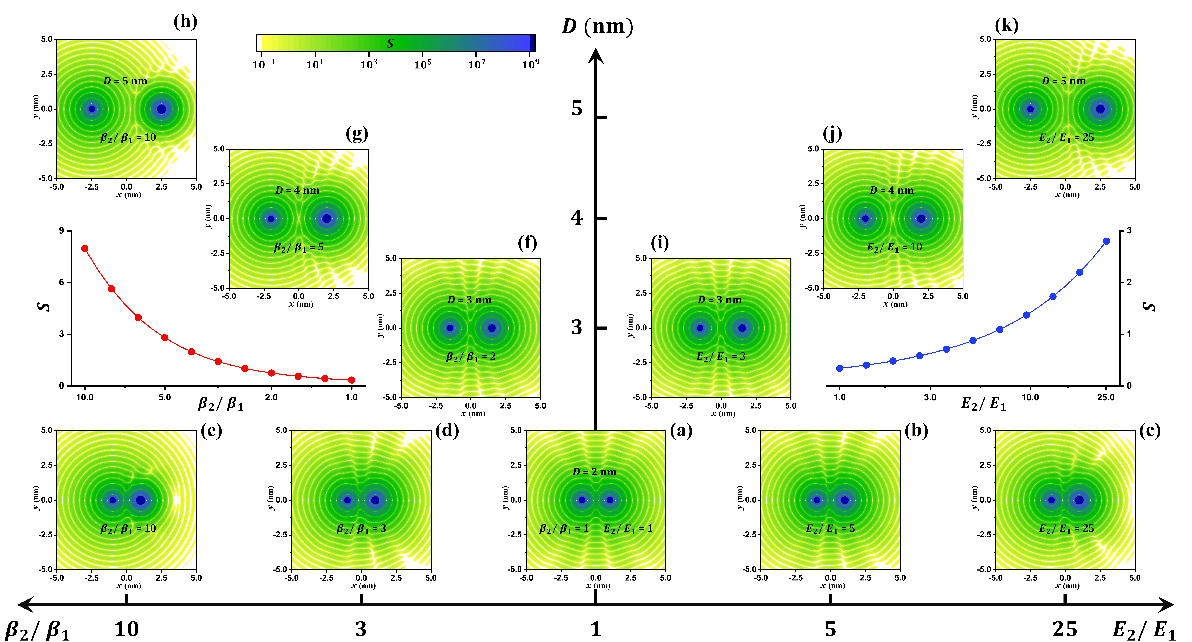}
\caption{The evolution of phonon coherent patterns induced by neutral defect pairs with the ratios of energy-level depth ($E_2/E_1$) and the localization strength parameter ($\beta_1/\beta_2$) for defect pairs and the spatial distance between two defects. Huang-Rhys factors that arising from the contribution of the selective phonon mode ($\omega_k=76$ THz) via the spatial integral around defect pairs as function of the ratio of $E_2/E_1$ and $\beta_1/\beta_2$ in the right-hand and left-hand insets, respectively. The energy-level depth $E_1$=0.1 eV among the bandgap for defect I is adopted.}
\end{figure*}

We illustrate the spatial evolution of lattice waves driven by positive-negative and neutral-neutral defects pairs in Fig. 1 (a) and (b), respectively. The oscillation amplitudes of them decay rapidly in the several nanometer scales with the very short period in tens of femtoseconds, which could be attributed to the local strength of charge states is reduced substantially away from the center of defect, resulting in the decrease of driving force arising from the electron-phonon coupling, and eventually the damping of lattice waves. This is also the primary obstacle for phonon interference in the current experimental measurements\cite{e14,e16,y7,y8}.

Under the assumption that the same frequency and the fixed difference of initial phase for two lattice waves, the interference patterns in several nanometers scale for two types of defect pairs are plotted in Fig.1 (c) and (d). Analogous to the interference effects of electrons and photons, the distinct interference fringes of lattice waves are appeared with the asymmetry and symmetry coherent patterns around the positive-negative and neutral-neutral defect pairs, respectively. More importantly, the intensities of constructive and destructive fringes are expressed in the units of Huang-Rhys factor. Namely, the displacive excitation of coherent lattice waves would be described by a measurable quantity in experiments\cite{h3,h4}, which will allow for the directly mapping interference patterns by optical spectroscopy as shown in the third section. The interference patterns show the different styles for two types of defect pairs. This can be attributed to the difference of internal charge states of defects pairs, because the amplitudes of lattice waves depend on the the distribution of local charge states as depicted in Eqs. (1) and (2), which is in good agreement with the previously theoretical prediction\cite{t1,t2,t3,t4,t5} and most experimental identifications\cite{e9,e14,e17}. In turn, some internal information for defect pairs could be extracted from styles of their interference phonon patterns. The phonon interference patterns for other seven charge states of defect pairs are shown in Fig. S3 of Supplemental Materials.

$\emph{The evolution of phonon interference patterns}.$-- Defect pair serving as the driving force for the coherent lattice dynamics, except for its charge state as mentioned above, other inherently internal parameters, including the spatial distance ($D$), the ratios of energy-depth ($E_1/E_2$) and localization strength ($\beta_1/\beta_2$) between two defects, should have substantial impacts on the interference processes of lattice waves. We illustrate the neural-neutral defect pair as an example to display the evolution of coherent patterns with these three internal parameters in Fig. 2.  Along the right-hand axis, one can see that the distribution of interference fringes evolute from the symmetry to the asymmetry patterns with increasing the ratio of energy-level depth of two defects, where the depth of defect (I) is fixed and defect (II) is increased. In general, the lattice distortion effect due to electron-phonon coupling are enhanced significantly as the energy-level depth of defect increases\cite{d1,h3,h4,h5}, which results in the unequally driven forces stemming from two defects for the coherent lattice waves, and thus the asymmetry interference patterns. The enhancement of lattice distortion could be further confirmed by the enlarged Huang-Rhys factor by the spatial integral of the interference fringes for the specific phonon mode around two defects as shown in the inset (blue dot-line). Interference patterns follow the similar trends for varying the ratio of localization strength along the left-hand axis, because the difference of lattice distortion effect between two defects could be modified substantially by this ratio parameter, also reflecting by the increasing of Huang-Rhys factor directly as shown in the inset (red dot-line). Here, the difference of localization strength between two neutral defects may be arising from (i) the discrepancy of the fundamental causes for defect formation\cite{h5,h6}, i.e. doping and dislocation; (ii) both of them could be modulated by the external perturbations\cite{t3,t4,t5}, such as infrared absorption, the electric field and so on.

Comparing with two ratio parameters, the spatial distance between two defects has more significant impact on the geometry of the interference pattern because the redistribution of the local charge-densities of defect pair are modified directly by this distance. In particular, the pattern of destructive fringes converted into the Dirac-like cones between two defects at certain values of the spatial distance cooperating with the appropriate ratios of energy-level depth or the localization strength as plotted in Fig. 2 (g) and (j), respectively. More interesting, the Dirac point evolutes into a line as these three parameters increase as given in Fig. 2 (h) and (k), respectively. These behaviors imply that the intriguing topology-like quantum states could be induced by phonon interference driven by electron-phonon coupling. In fact, a series of topological phenomena, i.e. ultrafast topological Lifshitz transition, the transient renormalization of topological band structure, have been proved by recently experiments\cite{e9,e10,e17}. However, the microscopic dynamical mechanisms for these topological states are still unclear. Obviously, the interference of lattice waves launched by defect pair sheds light on the underlying physics for these quantum states and offers an ideal platform to explore these ultrafast unequilibrium physics due to the electron-coherent phonon coupling. In addition, the evolution of interference phonon patterns for positive-negative defects pairs are also given in Fig. S4 in the Supplemental Materials.
\begin{figure}
\centering
\includegraphics[width=3.5in,keepaspectratio]{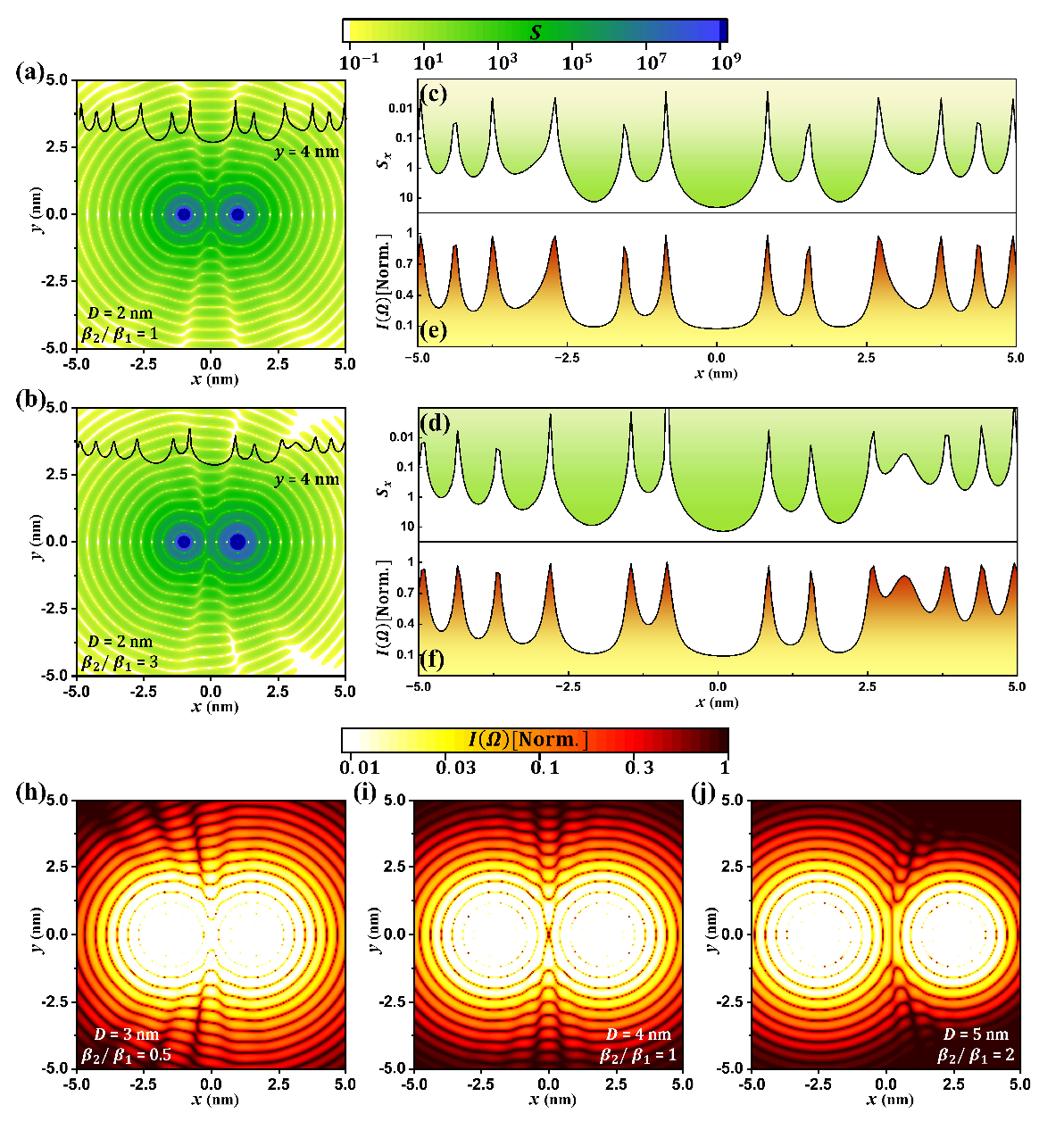}
\caption{The spatial evolution of interference intensities shown by green shadow lines in (c) and (d) along the $x$ axis extracted from symmetry (a) and asymmetry (b) coherent patterns induced by neutral defect pairs with the localization ratio parameter $\beta_1/\beta_2$ =1 and 3, respectively, at fixed energy-level depth, in which the corresponding simulated spanning spectroscopies are given as shown by red shadow lines in (e) and (f). (h)-(j) show the simulated spanning spectroscopies for interference patterns induced by defect pairs with four different localization ratios parameters within their coherent duration scales.}
\end{figure}

$\emph{The spectroscopic fingerprints}.$--
In fact, the identification of phonon interference is challenging because of its ultrafast duration ranging from picosecond to hundreds of femtosecond timescale. The well-established ultrafast spectroscopy technology\cite{e3,e4,e5,e5,e7,e11,e14,e16} cooperating with the measurable intensities of interference fringes reflected by Huang-Rhys factor, may provide the precursors for mapping the interference of lattice waves in the real space. On the other hand, defects in semiconductors serve as the color centers have been extensively proved due to their strong lattice relaxation effect\cite{h3,h4,h5}, where the average phonon number or \textit{phonon} $\textit{cloud}$ around defects are clearly reflected by the multiphonon replicas of optical spectroscopy and multiphonon Raman scattering\cite{r1,r2}. Next, we will propose a strategy for spanning phonon interference patterns by the ultrafast spectroscopy.

In the frame of Huang-Rhys model, the spectroscopic function for a defect pair can be given by\cite{h1,h2}
\begin{eqnarray}
I(\Omega)&\propto&\frac{2\pi M_{el}^2}{\hbar\Omega}(\frac{\overline{n}+1}{\overline{n}})^{p/2}e^{-S(2\overline{n}+1)}I_p(2S\sqrt{\overline{n}(\overline{n}+1)})\nonumber\\
&&\delta(\hbar\Omega-S\hbar\omega_k-p\hbar\omega_k),
\end{eqnarray}
where $M_{el}$ is the electron-photon interaction matrix, $\hbar\Omega$ is the energy of photon; $\overline{n}=[\exp{(\hbar\omega_k/k_BT)}-1]^{-1}$ denotes the thermal average number of the k$th$ phonon mode, $k_B$ is the Boltzmann constant; $I_p$ is the modified Bessel function of order $p$, with $p$ representing the number of absorbed or emitted phonons in the optical transition processes, $S$ is the Huang-Rhys factor calculated from Eq. (4) for the defect pair.

In Fig. 3 (c) and (d) present the evolution of interference intensities along $x$ axis at a fixed coordinate value of $y$, which are extracted from interference patterns for neutral defect pairs with the ratio parameter $\beta_1/\beta_2$ =1 and 3 as plotted in Fig. 3 (a) and (b), respectively. Meanwhile, the corresponding one dimensional emission spectroscopies are simulated for them based on Eq. (5) in their coherent duration shown in Fig. 3 (e) and (f). From the comparison, we find that the distribution of spectroscopic intensities are inverse proportion to Huang-Rhys factors, in very agreement with the characteristic features of spectroscopy for defects in most theoretical and experimental studies\cite{h4,h5,d3}, which results in the intensity of interference respecting to the amplitude of spectroscopy following the $\textit{lock-key}$ pattern or the $\textit{zipper}$ pattern. Fig. 3 (h)-(j) plot the simulated spanning optical spectroscopy for interference patterns induced by defect pairs with different localization ratios at room temperature, in which the outline of interference patterns, especially for the evolution of destructive fringes from Dirac conical to the nodal-line topological states between defect pairs could be mapped clearly.

$\emph{Discussion and conclusion}.$--
During the past several decades, in fact, the quantum beating signals have been observed by ultrafast spectroscopic techniques\cite{q1} in various materials systems, especially for the molecular and biological systems\cite{q2,q3,q4,q5}. Although several theoretical models\cite{q5} have assigned these behaviors to the vibronic mechanism arising from the coherent effect between electronic and vibrational motions, the exact nature of the vibronic system is not yet fully understood. The present model could provide fundamental insights for this mechanism from two aspects: (i) the oscillation of Huang-Rhys factor in the interference patterns originating from the coherent effect between the local charge states of defect pair and its surrounding lattice vibration, which, hence, may reveal the underlying nature of these quantum beats in physics; (ii) the vibronic mechanism proposed in most experiments and theories is based on coherent exciton-phonon coupling\cite{q3,q4,q5}, where the structural features of exciton, including the local charge states, the spatial distance of electron-hole and the ratio of effective masses between electron and hole, paly the predominate roles in determining the properties of quantum beats, which are very analogous to the evolution of phonon interference induced by positive-negative defect pair (see Fig. S4 in Supplemental Materials).

In addition, a series of low-energy quantum states around defect pair, except for the novel topology-like quantum states between two defects, are induced by the destructive interference fringes, representing low dissipation channels in certain directions and differing from the thermal equilibrium cases\cite{y6,y7,y8}, which allow for the ultrafast electron and energy transfer from the potential energy surface of one defect to another and would be manifested by the electron transport and scattering processes. But the measurement of thermal conductivity in practice will still very difficult due to the rapid decay of the interference pattern\cite{y8}.

In conclusion, we propose a microscopic model for the interference of lattice waves induced by defect pairs based on the well-known Huang-Rhys theory, where the lattice wave is driven independently by each of defects due to the local strong electron-phonon coupling effect. We present the spatial evolution of phonon interference around the defect pair and find that the intensities of  interference fringes could be directly evaluated by Huang-Rhys factors observed extensively in experiments. The evolution of geometrical patterns of phonon interference are determined by several internal parameters of defect pair, including the charge states, the spatial distance, the energy-level depth and the localization strength. In particular, the novel quantum states, such as the Dirac-like states are formed by destructive fringes for defect pairs with the appropriate internal parameters, shedding lights on the substantial modulation of topological states by the coherent phonon in recently experiments. Furthermore, we introduce the optical transition formula for defect pairs excited as the radiative centers based on the Huang-Rhys model and simulate their ultrafast optical spectroscopy within the phonon interference duration, where interference fringes are integrally reappeared by intensities of spectrum, just like the $\textit{lock-key}$ patterns, indicating some internal characteristics of defect pairs could be extracted from the optical spectroscopic fingerprints induced by their phonon interference coats.

This work was supported by National Natural Science Foundation of China (Grant No. 12634003 and 12574077) and Frontier Fundamental Research Program of Tianjin University (Grant No. 2025XJ21-0015).


\begin{thebibliography}{61}
\expandafter\if
x\csname natexlab\endcsname\relax\def\natexlab#1{#1}\fi
\expandafter\ifx\csname bibnamefont\endcsname\relax
  \def\bibnamefont#1{#1}\fi
\expandafter\ifx\csname bibfnamefont\endcsname\relax
  \def\bibfnamefont#1{#1}\fi
\expandafter\ifx\csname citenamefont\endcsname\relax
  \def\citenamefont#1{#1}\fi
\expandafter\ifx\csname url\endcsname\relax
  \def\url#1{\texttt{#1}}\fi
\expandafter\ifx\csname urlprefix\endcsname\relax\def\urlprefix{URL }\fi
\providecommand{\bibinfo}[2]{#2}
\providecommand{\eprint}[2][]{\url{#2}}
\bibitem[{\citenamefont{{Y. Dubi}}(2011)}]{y1}
\bibinfo{author}{\bibnamefont{{Y. Dubi and M. D. Ventra}}}, \bibinfo{title}{Colloquium: heat flow and thermoelectricity in atomic and molecular junctions}, \bibinfo{journal}{Rev. Mod. Phys.} \textbf{\bibinfo{volume}{83}}, \bibinfo{pages}{131} \bibinfo{year}{(2011)}.

\bibitem[{\citenamefont{{N. Li}}(2012)}]{y2}
\bibinfo{author}{\bibnamefont{{N. Li, J. Ren, L. Wang, G. Zhang, P. H$\ddot{a}$nggi, and B. Li}}}, \bibinfo{title}{Phononics: Manipulating heat flow with electronic analogs and beyond}, \bibinfo{journal}{Rev. Mod. Phys.} \textbf{\bibinfo{volume}{84}}, \bibinfo{pages}{1045} \bibinfo{year}{(2012)}.

\bibitem[{\citenamefont{{A. de la Torre}}(2012)}]{j1}
\bibinfo{author}{\bibnamefont{{A. de la Torre, D. M. Kennes, M. Claassen, S. Gerber, J. W. McIver, and M. A. Sentef}}}, \bibinfo{title}{Nonthermal pathways to ultrafast control in quantum materials}, \bibinfo{journal}{Rev. Mod. Phys.} \textbf{\bibinfo{volume}{93}}, \bibinfo{pages}{041002} \bibinfo{year}{(2021)}.

\bibitem[{\citenamefont{{G. Chen}}(2021)}]{y3}
\bibinfo{author}{\bibnamefont{{G. Chen}}}, \bibinfo{title}{Non-Fourier phonon heat conduction at the microscale and nanoscale}, \bibinfo{journal}{Nat. Rev. Phys.} \textbf{\bibinfo{volume}{3}}, \bibinfo{pages}{555} \bibinfo{year}{(2021)}.

\bibitem[{\citenamefont{{J. Chen}}(2022)}]{y4}
\bibinfo{author}{\bibnamefont{{J. Chen, X. Xu, J. Zhou, and B. Li}}}, \bibinfo{title}{Interfacial thermal resistance: Past, present, and future}, \bibinfo{journal}{Rev. Mod. Phys.} \textbf{\bibinfo{volume}{94}}, \bibinfo{pages}{025002} \bibinfo{year}{(2022)}.

\bibitem[{\citenamefont{{M. N. Luckyanova}}(2012)}]{y5}
\bibinfo{author}{\bibnamefont{{M. N. Luckyanova, J. Garg, K. Esfarjani, A. Jandl, M. T. Bulsara, A. J. Schmidt, A. J. Minnich, S. Chen, M. S. Dresselhaus, Z. Ren, E. A. Fitzgerald, and G. Chen}}}, \bibinfo{title}{Coherent phonon heat conduction in superlattices}, \bibinfo{journal}{Science} \textbf{\bibinfo{volume}{338}}, \bibinfo{pages}{936} \bibinfo{year}{(2012)}.

\bibitem[{\citenamefont{{M. Maldovan}}(2015)}]{y6}
\bibinfo{author}{\bibnamefont{{M. Maldovan}}}, \bibinfo{title}{Phonon wave interference and thermal bandgap materials}, \bibinfo{journal}{Nat. Mater.} \textbf{\bibinfo{volume}{14}}, \bibinfo{pages}{667} \bibinfo{year}{(2015)}.

\bibitem[{\citenamefont{{S. C. Yelishala}}(2025)}]{y7}
\bibinfo{author}{\bibnamefont{{S. C. Yelishala, Y. Zhu, P. M. Martinez, H. Chen, M. Habibi, G. Prampolini, J. C. Cuevas, W. Zhang, J. G. Vilhena, and L. Cui}}}, \bibinfo{title}{Phonon interference in single-molecule junctions}, \bibinfo{journal}{Nat. Mater.} \textbf{\bibinfo{volume}{24}}, \bibinfo{pages}{1258}  \bibinfo{year}{(2025)}.

\bibitem[{\citenamefont{{Jennifer C. Brookes}}(2011)}]{y8}
\bibinfo{author}{\bibnamefont{{E. Scheer, P. Leiderer }}}, \bibinfo{title}{Detecting interference of lattice vibrations}, \bibinfo{journal}{Nat. Mater.} \textbf{\bibinfo{volume}{24}}, \bibinfo{pages}{1164}  \bibinfo{year}{(2025)}.

\bibitem[{\citenamefont{{G. A. Garrett}}(1996)}]{j2}
\bibinfo{author}{\bibnamefont{{G. A. Garrett, T. F. Albrecht, J. F. Whitaker, and R. Merlin}}}, \bibinfo{title}{Coherent THz phonons driven by light pulses and the Sb problem: What is the mechanism?}, \bibinfo{journal}{Phys. Rev. Lett.} \textbf{\bibinfo{volume}{77}}, \bibinfo{pages}{3663} \bibinfo{year}{(1996)}.

\bibitem[{\citenamefont{{T. E. Stevens}}(2002)}]{e1}
\bibinfo{author}{\bibnamefont{{T. E. Stevens, J. Kuhl, and R. Merlin}}}, \bibinfo{title}{Coherent phonon generation and the two stimulated Raman tensors}, \bibinfo{journal}{Phys. Rev. B} \textbf{\bibinfo{volume}{65}}, \bibinfo{pages}{144304} \bibinfo{year}{(2002)}.

\bibitem[{\citenamefont{{M. Hase}}(2005)}]{e2}
\bibinfo{author}{\bibnamefont{{M. Hase, K. Ishioka, J. Demsar, K. Ushida, and M.
Kitajima}}}, \bibinfo{title}{Ultrafast dynamics of coherent optical phonons and nonequilibrium electrons in transition metals}, \bibinfo{journal}{Phys. Rev. B} \textbf{\bibinfo{volume}{71}}, \bibinfo{pages}{184301} \bibinfo{year}{(2005)}.

\bibitem[{\citenamefont{{Y.-S. Lim}}(2006)}]{e3}
\bibinfo{author}{\bibnamefont{{Y.-S. Lim, K.-J. Yee, J.-H. Kim, E. H. H$\acute{a}$roz, J. Shaver, J. Kono, S. K. Doorn, R. H. Hauge, and R. E. Smalley}}}, \bibinfo{title}{Coherent Lattice Vibrations in Single-Walled Carbon Nanotubes}, \bibinfo{journal}{Nano Lett.} \textbf{\bibinfo{volume}{6}}, \bibinfo{pages}{2696} \bibinfo{year}{(2006)}.

\bibitem[{\citenamefont{{M. F$\ddot{o}$rst}}(2011)}]{e4}
\bibinfo{author}{\bibnamefont{{M. F$\ddot{o}$rst, C. Manzoni, S. Kaiser, Y. Tomioka, Y. Tokura, R. Merlin, and A. Cavalleri}}}, \bibinfo{title}{ Nonlinear phononics as an ultrafast route to lattice control}, \bibinfo{journal}{Nat. Phys.} \textbf{\bibinfo{volume}{7}}, \bibinfo{pages}{854} \bibinfo{year}{(2011)}.

\bibitem[{\citenamefont{{E. Papalazarou}}(2012)}]{e5}
\bibinfo{author}{\bibnamefont{{E. Papalazarou, J. Faure, J. Mauchain, M. Marsi, A. Taleb-Ibrahimi, I. Reshetnyak, A. van Roekeghem, I. Timrov, N. Vast, B. Arnaud, and L. Perfetti}}}, \bibinfo{title}{Coherent phonon coupling to individual Bloch states in photoexcited bismuth}, \bibinfo{journal}{Phys. Rev. Lett.} \textbf{\bibinfo{volume}{108}}, \bibinfo{pages}{256808} \bibinfo{year}{(2012)}.

\bibitem[{\citenamefont{{E. Golias}}(2016)}]{e6}
\bibinfo{author}{\bibnamefont{{E. Golias and J. S$\acute{a}$nchez-Barriga}}}, \bibinfo{title}{Observation of antiphase coherent phonons in the warped Dirac cone of Bi$_2$Te$_3$}, \bibinfo{journal}{Phys. Rev. B} \textbf{\bibinfo{volume}{94}}, \bibinfo{pages}{161113(R)} \bibinfo{year}{(2016)}.

\bibitem[{\citenamefont{{D. M. Juraschek}}(2017)}]{e7}
\bibinfo{author}{\bibnamefont{{D. M. Juraschek, M. Fechner, and N. A. Spaldin}}}, \bibinfo{title}{Ultrafast Structure Switching through Nonlinear Phononics}, \bibinfo{journal}{Phys. Rev. Lett.} \textbf{\bibinfo{volume}{118}}, \bibinfo{pages}{054101}  \bibinfo{year}{(2017)}.

\bibitem[{\citenamefont{{D. M. Juraschek}}(2018)}]{e8}
\bibinfo{author}{\bibnamefont{{D. M. Juraschek and S. F. Maehrlein}}}, \bibinfo{title}{Sum-frequency ionic
Raman scattering}, \bibinfo{journal}{Phys. Rev. B} \textbf{\bibinfo{volume}{97}}, \bibinfo{pages}{174302} \bibinfo{year}{(2018)}.

\bibitem[{\citenamefont{{P. Hein}}(2020)}]{e9}
\bibinfo{author}{\bibnamefont{{P. Hein, S. Jauernik, H. Erk, L. Yang, Y. Qi, Y. Sun, C. Felser, and M. Bauer}}}, \bibinfo{title}{ Mode-resolved reciprocal space mapping of electron-phonon interaction in the Weyl semi-metal candidateTd-WTe$_2$}, \bibinfo{journal}{Nat. Commun.} \textbf{\bibinfo{volume}{11}}, \bibinfo{pages}{2613} \bibinfo{year}{(2020)}.

\bibitem[{\citenamefont{{C. Vaswani}}(2020)}]{e10}
\bibinfo{author}{\bibnamefont{{C. Vaswani, L.-L. Wang, D. H. Mudiyanselage, Q. Li, P. M. Lozano, G. D. Gu, D. Cheng, B. Song, L. Luo, R. H. J. Kim, C. Huang, Z. Liu, M. Mootz, I. E. Perakis, Y. Yao, K. M. Ho, and J. Wang}}}, \bibinfo{title}{Light-driven Raman coherence as a nonthermal route to ultrafast topology switching in a Dirac semimetal}, \bibinfo{journal}{Phys. Rev. X} \textbf{\bibinfo{volume}{10}}, \bibinfo{pages}{021013} \bibinfo{year}{(2020)}.

\bibitem[{\citenamefont{{D. Afanasiev}}(2021)}]{e11}
\bibinfo{author}{\bibnamefont{{D. Afanasiev, J. R. Hortensius, B. A. Ivanov, A. Sasani, E. Bousquet, Y. M. Blanter, R. V. Mikhaylovskiy, A. V. Kimel, and A. D. Caviglia}}}, \bibinfo{title}{Ultrafast control of magnetic interactions via light-driven phonons}, \bibinfo{journal}{Nat. Mater.} \textbf{\bibinfo{volume}{20}}, \bibinfo{pages}{607} \bibinfo{year}{(2021)}.

\bibitem[{\citenamefont{{A. S. Disa}}(2021)}]{e12}
\bibinfo{author}{\bibnamefont{{A. S. Disa, T. F. Nova, and A. Cavalleri}}}, \bibinfo{title}{Engineering crystal structures with light}, \bibinfo{journal}{Nat. Phys.} \textbf{\bibinfo{volume}{17}}, \bibinfo{pages}{1087} \bibinfo{year}{(2021)}.

\bibitem[{\citenamefont{{J. L. Bartolome}}(2022)}]{e13}
\bibinfo{author}{\bibnamefont{{J. L. Bartolome, C. Lian, W. H. Sio, I. G. Gurtubay, A. Eiguren, and F. Giustino}}}, \bibinfo{title}{Unified approach to polarons and phonon-induced band structure renormalization}, \bibinfo{journal}{Phys. Rev. Lett.} \textbf{\bibinfo{volume}{129}}, \bibinfo{pages}{076402} \bibinfo{year}{(2022)}.

\bibitem[{\citenamefont{{S. Liu}}(2022)}]{e14}
\bibinfo{author}{\bibnamefont{{S. Liu, A. Hammud, I. Hamada, M. Wolf, M. M$\ddot{u}$ller, T. Kumagai}}}, \bibinfo{title}{Nanoscale coherent phonon spectroscopy}, \bibinfo{journal}{Sci. Adv.} \textbf{\bibinfo{volume}{8}}, \bibinfo{pages}{eabq5682} \bibinfo{year}{(2022)}.

\bibitem[{\citenamefont{{C. Wang}}(2023)}]{e15}
\bibinfo{author}{\bibnamefont{{C. Wang, X. Liu, Q. Chen, D. Chen, Y. Wang, and S. Meng}}}, \bibinfo{title}{Coherent-Phonon-Driven Intervalley Scattering and Rabi Oscillation in Multivalley 2D Materials}, \bibinfo{journal}{Phys. Rev. Lett.} \textbf{\bibinfo{volume}{131}}, \bibinfo{pages}{066401} \bibinfo{year}{(2023)}.

\bibitem[{\citenamefont{{J. Zhang}}(2024)}]{e16}
\bibinfo{author}{\bibnamefont{{J. Zhang, Z. Wang, F. Lengers, D. Wigger, D. E. Reiter,T. Kuhn, H. J. W$\ddot{o}$rner, and T. T. Luu}}}, \bibinfo{title}{High-harmonic spectroscopy probes lattice dynamics}, \bibinfo{journal}{Nat. Photon.} \textbf{\bibinfo{volume}{18}}, \bibinfo{pages}{792} \bibinfo{year}{(2024)}.

\bibitem[{\citenamefont{{C. Emeis}}(2025)}]{e17}
\bibinfo{author}{\bibnamefont{{C. Emeis, S. Jauernik, S. Dahiya, Y. Pan, C. E. Jensen, P. Hein, M. Bauer, and F. Caruso}}}, \bibinfo{title}{Coherent Phonons and Quasiparticle Renormalization in Semimetals from First Principles}, \bibinfo{journal}{Phys. Rev. X} \textbf{\bibinfo{volume}{15}}, \bibinfo{pages}{021039} \bibinfo{year}{(2025)}.

\bibitem[{\citenamefont{{C. Wang}}(2025)}]{e18}
\bibinfo{author}{\bibnamefont{{C. Wang, D. Chen,Y. Wang, and S. Meng}}}, \bibinfo{title}{Directional Pumping of Coherent Phonons and Quasiparticle Renormalization in a Dirac Nodal-Line Semimetal}, \bibinfo{journal}{Phys. Rev. X} \textbf{\bibinfo{volume}{15}}, \bibinfo{pages}{021053} \bibinfo{year}{(2025)}.

\bibitem[{\citenamefont{{H. J. Zeiger}}(1992)}]{t1}
\bibinfo{author}{\bibnamefont{{H. J. Zeiger, J. Vidal, T. K. Cheng, E. P. Ippen, G. Dresselhaus, and M. S. Dresselhaus}}}, \bibinfo{title}{Theory for displacive excitation of coherent phonons}, \bibinfo{journal}{Phys. Rev. B} \textbf{\bibinfo{volume}{45}}, \bibinfo{pages}{768} \bibinfo{year}{(1992)}.

\bibitem[{\citenamefont{{A. V. Kuznetsov}}(1994)}]{t2}
\bibinfo{author}{\bibnamefont{{A. V. Kuznetsov and C. J. Stanton}}}, \bibinfo{title}{Theory of coherent
phonon oscillations in semiconductors}, \bibinfo{journal}{Phys. Rev. Lett.} \textbf{\bibinfo{volume}{73}}, \bibinfo{pages}{3243} \bibinfo{year}{(1994)}.

\bibitem[{\citenamefont{{M. Lakehal}}(2019)}]{t3}
\bibinfo{author}{\bibnamefont{{M. Lakehal and I. Paul}}}, \bibinfo{title}{Microscopic description of displacive coherent phonons}, \bibinfo{journal}{Phys. Rev. B} \textbf{\bibinfo{volume}{99}}, \bibinfo{pages}{035131} \bibinfo{year}{(2019)}.

\bibitem[{\citenamefont{{D. Kim}}(2022)}]{t4}
\bibinfo{author}{\bibnamefont{{D. Kim, A. Aydin, A. Daza, K. N. Avanaki, J. K. Rahkonen, and E. J. Heller}}}, \bibinfo{title}{Coherent charge carrier dynamics in the presence of thermal lattice vibrations}, \bibinfo{journal}{Phys. Rev. B} \textbf{\bibinfo{volume}{106}}, \bibinfo{pages}{054311} \bibinfo{year}{(2022)}.

\bibitem[{\citenamefont{{F. Caruso}}(2023)}]{t5}
\bibinfo{author}{\bibnamefont{{F. Caruso and M. Zacharias}}}, \bibinfo{title}{Quantum theory of light-driven
coherent lattice dynamics}, \bibinfo{journal}{Phys. Rev. B} \textbf{\bibinfo{volume}{107}}, \bibinfo{pages}{054102} \bibinfo{year}{(2023)}.

\bibitem[{\citenamefont{{F. Giustino}}(2017)}]{t6}
\bibinfo{author}{\bibnamefont{{F. Giustino}}}, \bibinfo{title}{Electron-phonon interactions from first principles}, \bibinfo{journal}{Rev. Mod. Phys.} \textbf{\bibinfo{volume}{89}}, \bibinfo{pages}{015003} \bibinfo{year}{(2017)}.

\bibitem[{\citenamefont{{K. Huang}}(1950)}]{h1}
\bibinfo{author}{\bibnamefont{{K. Huang and A. Rhys}}}, \bibinfo{title}{Theory of light absorption and non-radiative transitions in F-centres}, \bibinfo{journal}{Proc. R. Soc. A} \textbf{\bibinfo{volume}{204}}, \bibinfo{pages}{406} \bibinfo{year}{(1950)}.

\bibitem[{\citenamefont{{K. Huang}}(1981)}]{h2}
\bibinfo{author}{\bibnamefont{{K. Huang}}}, \bibinfo{title}{Adiabatic approximation theory and static coupling theory of nonradiative transition}, \bibinfo{journal}{Sci. Sinica} \textbf{\bibinfo{volume}{24}}, \bibinfo{pages}{27} (\bibinfo{year}{1981}).

\bibitem[{\citenamefont{{M. Hase}}(2010)}]{h3}
\bibinfo{author}{\bibnamefont{{M. Hase and M. Kitajima}}}, \bibinfo{title}{Interaction of coherent phonons with defects and elementary excitations}, \bibinfo{journal}{J. Phys. Condens. Matter} \textbf{\bibinfo{volume}{22}}, \bibinfo{pages}{073201} \bibinfo{year}{(2010)}.

\bibitem[{\citenamefont{{Y. Zhang}}(2019)}]{h4}
\bibinfo{author}{\bibnamefont{{Y. Zhang}}}, \bibinfo{title}{Applications of Huang-Rhys theory in semiconductor optical spectroscopy}, \bibinfo{journal}{J. Semicond.} \textbf{\bibinfo{volume}{40}}, \bibinfo{pages}{091102} (\bibinfo{year}{2019}).

\bibitem[{\citenamefont{{C. E. Dreyer}}(2024)}]{h5}
\bibinfo{author}{\bibnamefont{{C. E. Dreyer, A. Janotti, J. L. Lyons, and D. Wickramaratne}}}, \bibinfo{title}{Defects in semiconductors}, \bibinfo{journal}{J. Appl. Phys.} \textbf{\bibinfo{volume}{136}}, \bibinfo{pages}{190401} (\bibinfo{year}{2024}).

\bibitem[{\citenamefont{{C. E. Dreyer}}(2024)}]{h6}
\bibinfo{author}{\bibnamefont{{J. Klein, K. M. Roccapriore, M. Weile, S. Grytsiuk, A. R. Lupini, Z. Sofer, D. Pashov, M. van Schilfgaarde, S. Acharya, M. R$\ddot{o}$sner, and F. M. Ross}}}, \bibinfo{title}{Mesoscale atomic engineering in a crystal lattice}, \bibinfo{journal}{Nature} \textbf{\bibinfo{volume}{653}}, \bibinfo{pages}{715} (\bibinfo{year}{2026}).

\bibitem[{\citenamefont{{S. T. Pantelides}}(1978)}]{d1}
\bibinfo{author}{\bibnamefont{{S. T. Pantelides}}}, \bibinfo{title}{The electronic structure of impurities and other point defects in semiconductors}, \bibinfo{journal}{Rev. Mod. Phys.} \textbf{\bibinfo{volume}{50}}, \bibinfo{pages}{797} (\bibinfo{year}{1978}).

\bibitem[{\citenamefont{{D. Swiata}}(1991)}]{d2}
\bibinfo{author}{\bibnamefont{{D. Swiata and W. M. Bartczak}}}, \bibinfo{title}{Calculation of the Huang-Rhys factor for electron capture by a neutral impurity}, \bibinfo{journal}{ Phys. Rev. B} \textbf{\bibinfo{volume}{43}}, \bibinfo{pages}{6776} \bibinfo{year}{(1991)}.

\bibitem[{\citenamefont{{B. K. Ridley}}(2013)}]{d3}
\bibinfo{author}{\bibnamefont{{B. K. Ridley}}}, \bibinfo{titel}{$Quantum$ $Processes$ $in$ $Semiconductors$ (Oxford University Press, Oxford, 2013)}.

\bibitem[{\citenamefont{{E. P. Pokatilov}}(2002)}]{r1}
\bibinfo{author}{\bibnamefont{{E. P. Pokatilov, S. N. Klimin, V. M. Fomin, J. T. Devreese, and F. W. Wise}}}, \bibinfo{title}{ Multiphonon Raman scattering in semiconductor nanocrystals: Importance of nonadiabatic transitions}, \bibinfo{journal}{ Phys. Rev. B} \textbf{\bibinfo{volume}{65}}, \bibinfo{pages}{075316} \bibinfo{year}{(2002)}.

\bibitem[{\citenamefont{{Z.-W. Wang}}(2019)}]{r2}
\bibinfo{author}{\bibnamefont{{Z.-W. Wang, Y. Xiao, J.-P. Deng, Y. Cui , and Z.-Q. Li}}}, \bibinfo{title}{Multiphonon Raman scattering mediated by the exciton states in monolayer transition metal chalcogenides}, \bibinfo{journal}{ Phys. Rev. B} \textbf{\bibinfo{volume}{100}}, \bibinfo{pages}{125308} \bibinfo{year}{(2019)}.

\bibitem[{\citenamefont{{D. M. Jonas}}(2003)}]{q1}
\bibinfo{author}{\bibnamefont{{D. M. Jonas}}}, \bibinfo{title}{Two- dimensional femtosecond spectroscopy}, \bibinfo{journal}{Annu. Rev. Phys. Chem} \textbf{\bibinfo{volume}{54}}, \bibinfo{pages}{425} \bibinfo{year}{(2003)}.

\bibitem[{\citenamefont{{G. S. Engel}}(2007)}]{q2}
\bibinfo{author}{\bibnamefont{{G. S. Engel, T. R. Calhoun, E. L. Read, T.-K. Ahn, T. Manal, Y.-C. Cheng, R. E. Blankenship and G. R. Fleming}}}, \bibinfo{title}{Evidence for wavelike energy transfer through quantum coherence in photosynthetic systems}, \bibinfo{journal}{Nature} \textbf{\bibinfo{volume}{446}}, \bibinfo{pages}{782} \bibinfo{year}{(2007)}.

\bibitem[{\citenamefont{{A. Halpin}}(2014)}]{q3}
\bibinfo{author}{\bibnamefont{{A. Halpin, P. J. M. Johnson, R. Tempelaar, R. S. Murphy, J. Knoester, T. L. C. Jansen and R. J. D. Miller}}}, \bibinfo{title}{Two-dimensional spectroscopy of a molecular dimer unveils the effects of vibronic coupling on exciton coherences}, \bibinfo{journal}{Nat. Chem.} \textbf{\bibinfo{volume}{6}}, \bibinfo{pages}{196} \bibinfo{year}{(2014)}.

\bibitem[{\citenamefont{{G. D. Scholes}}(2017)}]{q4}
\bibinfo{author}{\bibnamefont{{G. D. Scholes, G. R. Fleming, L. X. Chen, A. A.-Guzik, A. Buchleitner, D. F. Coker, G. S. Engel, R. van Grondelle, A. Ishizaki, D. M. Jonas, J. S. Lundeen, J. K. McCusker, S. Mukamel, J. P. Ogilvie, A. Olaya-Castro, M. A. Ratner, F. C. Spano, K. B. Whaley and X. Zhu}}}, \bibinfo{title}{Using coherence to enhance function in chemical and biophysical systems}, \bibinfo{journal}{Nature} \textbf{\bibinfo{volume}{543}}, \bibinfo{pages}{647} \bibinfo{year}{(2017)}.

\bibitem[{\citenamefont{{L.Wang}}(2019)}]{q5}
\bibinfo{author}{\bibnamefont{{L.Wang, M. A. Allodi and G. S. Engel}}}, \bibinfo{title}{Quantum coherences reveal excited- state dynamics in biophysical systems}, \bibinfo{journal}{Nat. Rev. Chem.} \textbf{\bibinfo{volume}{3}}, \bibinfo{pages}{477} \bibinfo{year}{(2019)}.

\end{thebibliography}
\end{document}